# Predict Click-Through Rates with Deep Interest Network Model in E-commerce Advertising

Chang Zhou[1]
Columbia University
New York, USA
mmchang042929@gmail.com

Yang Zhao[2,*]
Columbia University
New York, USA
* Corresponding author: yangzhaozyang@gmail.com

Yuelin Zou[2]
Columbia University
New York, USA
yz4198@columbia.edu

Jin Cao[2]
Independent Researcher
Dallas, USA
caojinscholar@gmail.com

Wenhan Fan[3]
Independent Researcher
New York, USA
finncontactplus@gmail.com

Yi Zhao[4]
Independent Researcher
Sunnyvale, USA
zhaoyizjuee@gmail.com

Chiyu Cheng[5]
University of California, Irvine
Seattle, USA
cypersonal6@gmail.com

*Abstract*—This paper proposes new methods to enhance click-through rate (CTR) prediction models using the Deep Interest Network (DIN) model, specifically applied to the advertising system of Alibaba's Taobao platform. Unlike traditional deep learning approaches, this research focuses on localized user behavior activation for tailored ad targeting by leveraging extensive user behavior data. Compared to traditional models, this method demonstrates superior ability to handle diverse and dynamic user data, thereby improving the efficiency of ad systems and increasing revenue.

## I. Introduction

The e-commerce advertising system relies heavily on accurately predicting user clicks on displayed ads, a metric quantified by the CTR. Traditional models often fail to account for the nuanced interests of individual users due to their simplistic handling of diverse and dynamic user data. In this paper, we introduce the Deep Interest Network (DIN) to address these shortcomings by focusing on localized user interests derived from historical behavior data, resulting in more accurate ad click predictions. Our proposed methods significantly enhance ad targeting precision and accuracy, thereby improving the economic efficiency of online advertising systems.

### A. Overview of Advertising Systems

When a user visits an e-commerce site, the ads system performs these following steps: retrieve user historical behavior; generate candidate ads; predicts click probabilities, and logs user reactions. This forms a closed-loop feedback system essential for refining future ad predictions.

### B. CTR Models

CTR prediction models are crucial for optimizing ad placement, affecting the system's revenue directly. Effective models can enhance user engagement by presenting ads that are more likely to be of interest, thereby improving both user experience and advertising efficiency.

- *CTR (Click-Through Rate):* CTR = Number of Impressions / Number of Clicks
- *eCPM (Effective Cost Per Mille):* eCPM = CTR $\times$ Bid Price

## II. Literature Review

The landscape of click-through rate (CTR) prediction has been primarily dominated by basic machine learning techniques, such as multilayer perceptrons (MLPs) and logistic regression. Historically, these methods have focused predominantly on static user profiles and have leveraged sparse feature interactions. Such traditional models have been critiqued for their inadequate capacity to unravel and represent the intricate structures embedded within behavioral data, often relying excessively on feature engineering and operating under oversimplified assumptions about user interactions. This reliance limits their effectiveness in addressing the nuanced nature of user behavior in digital environments.

The advent of deep learning[1-5] has heralded significant advancements in the field of CTR predictions. The application of MLPs, in particular, has been transformative, enabling the automatic learning of non-linear feature interactions through deep neural networks. This evolution has facilitated a more sophisticated understanding of input features, albeit with certain limitations. Notably, while deep learning models[6-11] excel at identifying complex patterns in large datasets, they often fall short of fully capturing the idiosyncratic and evolving nature of user behavior data specific to online advertising contexts. These models, though powerful, typically overlook the temporal dynamics that characterize user interactions, which are crucial for predicting future actions based on past behavior.

Moreover, despite recent strides in incorporating more advanced deep learning techniques[12-15] that integrate aspects of temporal dynamics — such as recurrent neural networks (RNNs) and their variants — these approaches still tend to neglect the individual variability in how user interests are triggered and evolve over time. There remains a significant gap in the ability of existing models to adaptively and dynamically synthesize disparate user data points for real-time predictive analysis.

This research seeks to address these deficiencies by exploring innovative modeling techniques that not only embrace the temporal and evolving nature of user data but also adaptively tune to the individual's fluctuating interests. Historical literature highlights a critical need for models that can dynamically integrate and analyze real-time data to make accurate predictions, thereby significantly enhancing the effectiveness of online advertising strategies. By filling this gap, we aim to contribute to the body of knowledge in computational advertising and offer practical solutions that can be deployed in dynamic, real-world environments, with improving the prediction accuracy on their user responses.

III. THEORETICAL FRAMEWORK

The theoretical underpinning of this research is based on the principles of machine learning[16-18] and deep learning[19-21], particularly focusing on representation learning and feature interactions within user behavior data.

A. Deep Interest Network (DIN)

DIN improves upon traditional models by using an interest distribution mechanism to represent diverse user interests and applying an attention-like network structure. This allows the model to focus on the most relevant user interests for each candidate ad.

B. Model Architecture

The DIN model architecture includes a base model where sparse ID features are transformed into dense embeddings, followed by MLPs to predict the CTR. The key innovation in DIN lies in its use of an *attention* mechanism, which adjusts the influence of various user behaviors dynamically based on the ad in question.

- *Interaction function:* $F(U, A) = V_\mu \cdot V_\alpha$ (1)
- *Weighted sum pooling in DIN:* $V_\mu = \sum_{i=1}^{N} \omega_i \cdot V_i$ (2)
- *Attention mechanism:* $\omega_i = g(V_i \cdot V_\alpha)$ (3)

Where $V_\mu$ is the user embedding, $V_\alpha$ is the ad embedding, $V_i$ are the embeddings of behavior IDs, and g is an activation function (e.g. softmax).

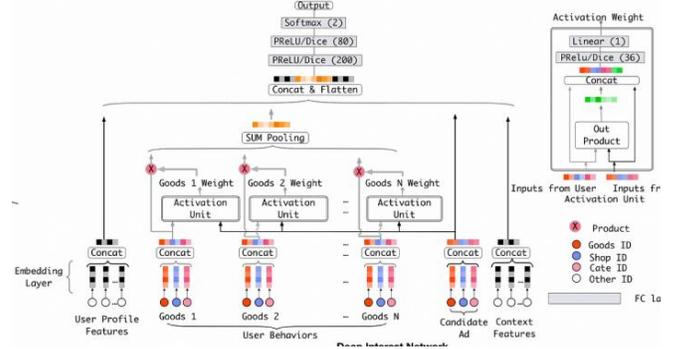

Figure 1. Deep interest network architecture

C. Implementation Details

In our research, we utilized the Alibaba X-Deep Learning (XDL) platform to implement the Deep Interest Network (DIN). The XDL platform is specifically engineered to optimize the training of deep learning models used in industrial-scale systems like the model training in e-commerce, where handling massive datasets efficiently is crucial.

*1) Platform Capabilities:* XDL is distinctively designed to support both model-parallelism and data-parallelism, which are crucial for the scalable training of complex models that require handling extensive amounts of data.

*a) Model-Parallelism:* This feature of XDL allows the distribution of different parts of the neural network model across multiple computing nodes. This is particularly useful for large models whose parameters cannot be efficiently handled by a single node. In the case of DIN, the model's embedding layers, which process sparse features, are distributed across various nodes. This distribution helps in managing the vast amount of sparse input data, especially involving in user interaction histories in e-commerce platforms.

*b) Data-Parallelism:* XDL also implements data-parallelism, where the training data gets split across different nodes. Each node processes a portion of the data with a complete model but shares parameters during training. This approach is effective in parallelizing the computational workload, thereby speeding up the training process significantly. Data-parallelism in XDL is facilitated by sophisticated synchronization mechanisms that ensure the consistency of model updates across all nodes, leading to efficient and accurate model training.

*2) Training Optimization:* Utilizing XDL's robust infrastructure, we were able to significantly enhance the training speed of the DIN model without compromising on its performance. The platform's ability to handle large-scale data and its sophisticated parameter update mechanisms allowed us to train our model with high efficiency, leveraging the distributed computing environment to reduce training time while maintaining high data throughput.

*3) Integration with E-commerce Systems:* The implementation of DIN on the XDL platform has been specifically optimized for integration with Alibaba's e-commerce systems. This integration allows for real-time data processing and immediate application of the trained model to predict user CTR on ads, demonstrating the practical utility and real-world applicability of our research.

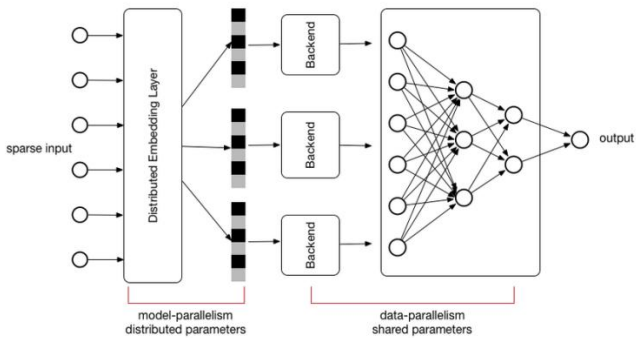

Figure 2. Architecture of XDL platform

## IV. METHODOLOGY

The study utilizes a robust dataset from Taobao, including user profiles, ad features, and raw ad display/click logs. The DIN model was implemented using the DeepCTR library, which is designed for deep learning-based CTR prediction tasks. Key features include:

*1) Data Preparation:* The raw data was meticulously preprocessed to transform categorical variables into a machine-readable format using one-hot encoding and to scale numerical values to ensure uniformity in data magnitude which aids in the optimization process.

*2) Model Construction:* The core of our methodology is the Deep Interest Network (DIN) model, implemented using the DeepCTR library, renowned for its efficacy in deep learning tasks targeting CTR prediction. The model architecture is particularly noted for its embedding layer that efficiently handles sparse features and an attention mechanism that dynamically focuses on the pertinent aspects of user behaviors with respect to each displayed ad.

*3) Training Protocol:* Training was conducted using the Adam optimizer. We employed binary cross-entropy loss as our loss function, complemented by regularization techniques to curb overfitting. This approach ensures that our model generalizes well on unseen data while maintaining robustness during training.

*4) Performance Evaluation:* Performance metrics such as AUC, log loss, and accuracy serve as evaluation criteria.

## V. EXPERIMENT

### A. Dataset

We utilized a non-public dataset, comprising 26 million ad display/click records over eight days. The dataset includes user demographics, ad characteristics, and user behavior patterns.

### B. Evaluation Metrics

To thoroughly assess the performance of our model, we implemented several key evaluation metrics:

*1) Group Weighted AUC (GAUC):* Our primary metric, the GAUC, offers a refined measure of model accuracy. Unlike traditional AUC, which treats all user interactions equally, GAUC allows for the weighting of user-specific scores based on their activity levels, such as the number of impressions or clicks each user has generated. This method provides a more nuanced and granular evaluation by acknowledging the varying influence of different user groups on the overall model performance. GAUC (Group weighted AUC) Calculation:

$$\text{GAUC} = \frac{\sum_{i=1}^{n} \omega_i \times \text{AUC}_i}{\sum_{i=1}^{n} \omega_i} \quad (4)$$

Where $\omega_i$ represents the weight for the i-th group, which could be the number of impressions or clicks.

GAUC Calculation Methodology: To compute the GAUC, we first assign weights to each user group based on their respective contribution to the dataset, such as the total number of impressions or clicks they have generated. These weights are then used to calculate the weighted average of the AUC scores for each group, ensuring that the contributions of more active users are appropriately emphasized in the overall assessment of the model.

*2) Log Loss:* As a secondary metric, log loss was utilized to measure the accuracy of the classifier. A lower log loss value indicates better predictions and a model's proficiency in estimating the probabilities of user clicks accurately.

*3) Accuracy:* Additionally, we measured the overall accuracy of the model in correctly predicting clicks versus non-clicks. This metric serves as a straightforward indicator of the model's effectiveness in classifying user responses.

## VI. FINDING AND DISCUSSION

The results indicate that the DIN model outperforms traditional models in terms of accuracy and log loss, suggesting a better handling of complex user behavior patterns and interest dynamics. The findings are discussed in the context of the theoretical framework, emphasizing how the attention mechanism provides a nuanced understanding of user interests, leading to more precise ad targeting.

The model's performance was evaluated using metrics like log loss and AUC (Area Under the Curve):

- *Log Loss:* The DIN model demonstrates a marked improvement in prediction accuracy, as reflected by a lower log loss which reached a nadir of approximately 0.08. This indicates fewer prediction errors and a higher reliability in forecasting user clicks.
- *Validation Loss and GAUC:* The DIN model not only shows a steady decline in validation loss but also an impressive rise in GAUC, peaking at around 0.75. These metrics illustrate the model's enhanced capability to differentiate between clicked and unclicked ads, thereby affirming its effectiveness in a real-world advertising scenario.

Theoretical and Practical Implications: The findings underscore the efficacy of integrating an attention mechanism within the DIN framework, which adaptively captures and utilizes complex user behaviors. This not only supports the theoretical model of dynamic user interest but also translates into more accurate ad targeting, potentially leading to higher user engagement and satisfaction on platforms like Taobao.

These insights are integral to advancing the field of computational advertising by emphasizing the role of tailored, behavior-focused algorithms in enhancing the relevance and efficiency of ad placements.

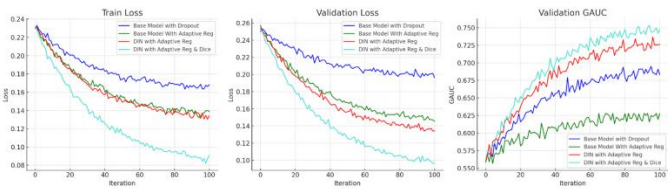

Figure 3.  Performance compares

## VII. CONCLUSION

This study proposes the new methods with implementation of the DIN model to enhance the performance and accuracy on the prediction of user clicks on ads using the real-world user behavior data, with further potential implications for enhancing the economic efficiency of advertising systems. Such method is evident from the improved log loss and AUC scores compared with traditional MLP models. In the future research, considering the complexity of model and high-demand computational requests, we could further explore the balance the scalability and efficiency of the models, with their applications across different platforms and ad types.